\documentclass{jltp}

\usepackage{graphicx}

\DeclareTextSymbol{\degre}{OT1}{23}

\title{Andreev Bound States and Josephson Coupling in  YBa$_2$Cu$_3$O$_{7-\delta}$/Pb tunnel junctions}

\author{J\'er\^ome Lesueur, Xavier Grison, Marco Aprili and Takis Kontos}

\address{CSNSM,  b\^atiment 108, Universit\'e Paris-Sud,
F-91405 ORSAY Cedex, France}

\runninghead{J.~Lesueur {\it et al.}}{ABS and Josephson Coupling in  YBCO/Pb tunnel junctions}

\begin{document}

\maketitle

\begin{abstract}

We have made in-situ [103] and [110]YBCO/Pb junctions, 
which display an hysteretic behaviour with both
a well-defined Quasi-Particule branch and a Josephson current. 
A zero energy conductance peak is observed in the tunneling conductance which is a signature of Andreev Bound States (ABS) at a surface of a $d$-wave superconductor. The observed splitting of this peak at low temperature may confirm the existence of a surface state which breaks time-reversal symmetry. 
 However, up to now,  we have not seen any
evidence of a coupling between Josephson current and the 
zero-energy ABS, predicted by Y.~Tanaka and S.~Kashiwaya\cite{tanaka}.


\end{abstract}

\section{INTRODUCTION}

The superconducting order parameter (OP) in High-T$_c$ cuprates, like YBa$_2$Cu$_3$O$_{7-\delta}$ (YBCO) is known to be mainly of $d_{x^2-y^2}$ symmetry in the bulk. Due to the sign change of the OP, Andreev Bound States develop on [110] surfaces\cite{hu}. Those ABS carry current at zero energy and therefore should couple to the Josephson current in Superconductor-Insulator-Superconductor (SIS) junctions, leading to a variety of novel and interesting features predicted by theories\cite{tanaka,fogelstrom}, but never observed experimentally to the best of our knowledge. We report on studies on YBCO/Pb tunnel junctions, made {\it in-situ}  to avoid interface contamination\cite{jlrc}. All the junctions display a clear hysteretic behavior (inset fig.\ref{fig1}), with a well defined Quasi-Particle (QP) branch and a Josephson current I$_c$. We report here on junctions along the [110] and [103] directions, the later being almost equivalent to the [100] one.

\section{EXPERIMENTAL RESULTS}

\subsection{Junctions fabrication}
YBCO films are grown by a reactive codeposition technique. Metallic vapors are condensed on a [110]~SrTiO$_3$ single crystal held at high temperature in an atmosphere of oxygen ($10^{-3}$ torrs), to make a 75~nm thick film in a rectangular shape. The optimal substrate temperature is 750\degre$\!$C to grow [103] oriented films. A 15~nm thick PrBa$_2$Cu$_3$O$_7$ template layer is deposited at 630\degre$\!$C, prior to the YBCO layer made at 750\degre$\!$C to grow [110] oriented films. The deposition chamber is then filled with pure oxygen while the temperature is decreased to room-temperature. Finaly, after vaccum it again, 60~nm thick Ag contact pads and 250~nm thick Pb counter-electrodes are cross deposited through shadow masks on top of the YBCO film. This {\it in-situ}  process leads to surface resistances 2 or 3 orders of magnitude lower than the usual {\it ex-situ}  ones, allowing to observe a sizable Josephson coupling at low temperature in these junctions.

\begin{figure}[btp]
\begin{center}\leavevmode
\includegraphics[width=0.9\linewidth]{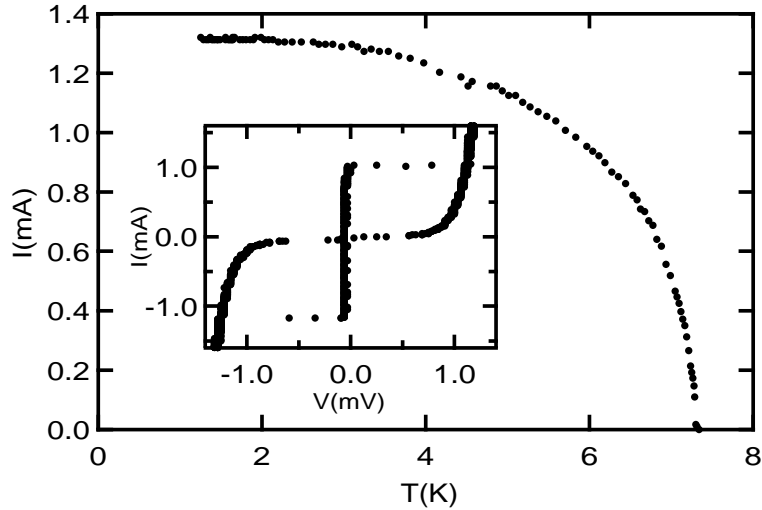}
\caption{ 
Temperature dependence of the critical current of a [103]YBCO/Pb junction. Inset: I(V) curve at 1.3~K.
}\label{fig1}\end{center}\end{figure}

\subsection{[103]YBCO/Pb junctions}

\begin{figure}[btp]
\begin{center}\leavevmode
\includegraphics[width=0.95\linewidth]{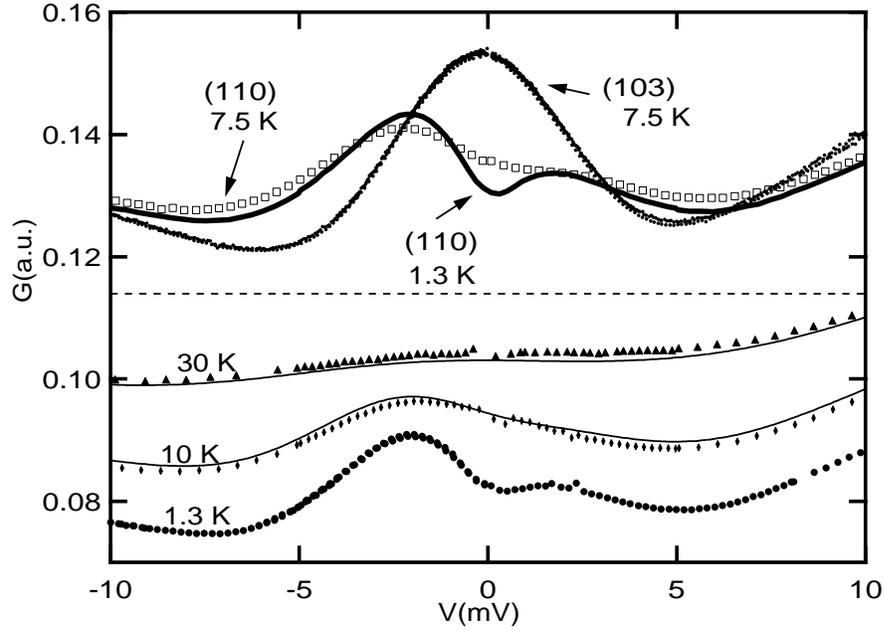}
\caption{ 
Upper part: Tunneling spectra of [110] and [103] junctions. Both display a ZBCP, split in the [110] case. Lower part: Temperature evolution of the split ZBCP of a [110]YBCO/Pb junction (curves are shifted for clarity). Symbols are experimental data. Lines are the convolution of the experimental data at 1.3~K with the derivative of the Fermi function.
Data at 1.3~K are taken under B=2500~G to suppress Pb superconductivity (At $T>T_{c_{Pb}}$, this field does not affect the tunneling spectra).
}\label{fig2}\end{center}\end{figure}

A reproducible Josephson coupling strength I$_c$R$_n \approx 200 \pm 50 \mu$V is observed, together with a magnetic field modulation of the critical current. The tunneling spectra exhibit a Zero Bias Conductance Peak (ZBCP), as high as 40\% of the background conductance, which is a signature of the ABS in the QP Density Of States (DOS)\cite{hu}. The occurrence of the ABS along this direction can be explained by a faceting of the interface in this geometry\cite{fogelstrom}. Since ABS and a Josephson current are observed simultaneously, we expect high order effects to take place\cite{tanaka}, and an unusual temperature dependence of the critical current to be seen, namely an upward curvature of the I$_c$(T) curve. This relies on the fact that both channels can carry current at zero energy, and that the ABS amplitude decreases roughly as $1/T$. Instead, a conventional square-root like behavior is observed (fig.\ref{fig1}), as if no coupling between the ABS and the Josephson current exists. Besides the fact that the above mentionned computation do not rely on self consistent calculations, two reasons may explain this experimental result. First, the film surface is faceted as observed by Atomic Force Microscopy (AFM) ; the ABS and the Josephson current may then come from different facets. Second, the amplitude of the ABS is small as to compare with those used in calculations, and therefore, could not contribute significantly to the current at low temperature.

\subsection{[110]YBCO/Pb junctions}

Together with the formation of the ABS, a strong depairing occurs in the [110] direction for a d$_{x^2-y^2}$ symmetry superconductor (as opposed to the [103] case), which may allow a subdominant component of the order parameter to rise up at the surface\cite{fogelstrom}, such as an $s$-wave one\cite{shiba}. In this case, a $d+is$ state forms within a few coherence length near the surface, below a transition temperature\cite{fogelstrom} T$^*$. This state breaks time-reversal symmetry ($\mathcal{T}$) since spontaneous supercurrents are generated to account for the dephasing between the two components. This yields to a Doppler-like shift of the ABS towards finite energies, and thus a splitting of the ZBCP\cite{covington}. We do observe a splitting $2 \delta \approx 4$mV of the ZBCP in our junctions (fig.\ref{fig2}), independently of the barrier resistance (from $10^{-5}\,\Omega\!\cdot\!\mbox{cm}^2$ to $10^{-2}\,\Omega\!\cdot\!\mbox{cm}^2$). A simple thermal smearing of the low temperature data (1.3K) can account for the high temperature one up to 30K, as expected for a DOS effect (fig.\ref{fig2}). We thus may conclude that surprisingly $T^* > 30$K in these junctions. This would give a $s$-wave component as high as 30 to 40\% of the $d$-wave one\cite{fogelstrom}, much higher than the reported one\cite{kouznetsov}. However, since the shift of the ABS relies on the depression of the dominant $d$-wave component of the OP at a surface, one should certainly take into account the role of the surface disorder to describe the details of this effect. A striking feature is the asymmetry of the peaks, in the opposite way of the background, which is not expected for split ABS, unless particle-hole symmetry is broken. This is robust against changes in the junction resistance (over 3~decades) and temperature variations. This gives us confidence that it is not barrier related, but a QP-DOS effect instead. Further work is needed to investigate this point.
These junctions also exhibit a Josephson current below T$_{c_{Pb}}$, and a hysteretic behavior. Once again, no clear signature of the coupling between the ABS and the Josephson current is observed. The Fraunhofer pattern of such junctions is not standard. However, one has to take into account the presence of the spontaneous currents along the interface\cite{stefanakis}.

\section{CONCLUSION}

We have made [110] and [103]YBCO/Pb junctions. Both show a Josephson current and ABS, but we have not yet seen any evidence of a coupling between them. In the [110] case, the ABS are split by surface-induced broken-$\mathcal{T}$ state. This splitting is asymmetric, what remains to be explained.

\section*{ACKNOWLEDGMENTS}

Authors thank F.~Lalu for his experimental support, G.~Passerieux, S.~J.~Kim and J.~Ayache for the microstructural characterization of the films.

\end{document}